\documentclass[final]{aipproc}

\layoutstyle{8x11double}

\usepackage{axodraw}
\usepackage{subfigure}

\def\eq{\begin{equation}}
\def\qe{\end{equation}}
\def\eqa{\begin{eqnarray}}
\def\qea{\end{eqnarray}}
\def\bib{\bibitem}

\newcommand{\newc}{\newcommand}
\newc{\softsusy}{\texttt{SOFTSUSY}}
\newc{\looptools}{\texttt{LoopTools}}
\newc{\isajet}{\texttt{ISAJET7.64}}

\newc{\RPVSUGRA}{RPV mSUGRA}

\newc{\MO}{m_0}
\newc{\Mhalf}{M_{1/2}}
\newc{\AO}{A_0}
\newc{\tanb}{\textrm{tan}\beta}
\newc{\sgnmu}{\textrm{sgn}{\mu}}

\newc{\lam}{\lambda}

\newc{\bsmumu}{BR(B_s\to \mu^+\mu^-)}
\newc{\bsgamma}{BR(b\to s \gamma)}
\newc{\MX}{M_X}
\newc{\MZ}{M_Z}

\newc{\lag}{\mathcal{L}}
\newc{\mM}{\mathcal{M}}
\newc{\mnu}{\m^{\nu}}

\begin{document}

\title{Neutrino masses in lepton number violating mSUGRA}


\author{Steve~C.~H.~Kom}{
  address={DAMTP, Centre for Mathematical Sciences, Wilberforce Road, Cambridge CB3 0WA, UK}
}

\begin{abstract}
In SUSY models which violate R-parity, there exist trilinear lepton number violating (LNV) operators which can lead to neutrino masses.  If these operators are defined at the unification scale, the renormalization group flow becomes important and generally leads to one neutrino mass much heavier than the others.  We study, in a minimal supergravity (mSUGRA) set-up with two trilinear LNV operators and three charged lepton mixing angles, numerically how these parameters may be arranged to be compatible with neutrino oscillation data, and discuss some phenomenological observations.
\end{abstract}

\maketitle

%%%%%%%%%%%%%%%%%%%%%%%%%%%%%%%%%%%%%%%%%%%%
%% MAINMATTER
%%%%%%%%%%%%%%%%%%%%%%%%%%%%%%%%%%%%%%%%%%%%

\section{mSUGRA models with lepton number violation}
In the minimal supersymmetric extension to the Standard Model (SM), a general superpotential contains simultaneously lepton and baryon number violating terms \cite{ssm-superpot}.  Their presence leads to fast proton decay \cite{proton-decay}, so at least one of the above set of terms need to be forbidden, for instance by some discrete symmetry \cite{Dreiner:2005rd}.  The most widely studied symmetry is R-parity \cite{Farrar:1978xj}, for which both sets of terms are projected out.  An alternative, known as baryon triality \cite{Ibanez:1991hv}, also suppresses fast proton decay but allows the set of lepton number violating (LNV) operators.  The presence of these LNV terms in turn leads to interesting phenomenological consequences such as neutrino masses and neutrinoless double beta decay.  It also opens up a large region of parameter space not available in R-parity conserving models, because the lightest supersymmetric particle (LSP) in the latter needs to be electrically and color neutral to avoid cosmological problems \cite{Ellis:1983ew}.  For these reasons, it is important to study models with LNV.

The renormalisable LNV superpotential is given by
\eqa \label{eq:superpot}
\mathcal{W}_{LNV}&=&\frac{1}{2}\lam_{ijk}L_iL_j\bar{E_k}
+\lam'_{ijk}L_iQ_j\bar{D_k} -\mu_iL_iH_u.
\qea
We shall focus on the dimensionless, trilinear operators $\lam_{ijk}$ and $\lam'_{ijk}$.  The supersymmetry (SUSY) breaking sector is specified using the minimal supergravity (mSUGRA) \cite{msugra} assumption.  We will attempt to account for the neutrino oscillation data \cite{0704.1800} (see Table \ref{tab:neutrino_data}) by two LNV parameters.  They are defined in a weak interaction basis in which the charged lepton Yukawa matrix $Y_E$ is symmetric but not diagonal, and the rotation to the diagonal basis can be characterized by three angles.  The parameter set that defines our model is therefore
\eqa \label{eq:par_set}
\MO, \Mhalf, \AO, \sgnmu,  &@& \MX, \nonumber \\
\theta^{l}_{12},\theta^{l}_{13},\theta^{l}_{23}, \Lambda_1, \Lambda_2 &@& \MX,\nonumber \\
 \tanb &@& \MZ, \qea
where $\Lambda_1,\Lambda_2 \in \{\lambda_{ijk},\lambda'_{ijk}\}$, and $\MX$ is the scale at which the electroweak gauge couplings unify.  $\MO$, $\Mhalf$ and $\AO$ are the universal scalar mass, gaugino mass and SUSY breaking trilinear scalar coupling at $\MX$ respectively, and $\sgnmu$ is the sign of $\mu$.  The charged lepton mixing angles $\theta^{l}$'s are defined in the standard parameterization \cite{PDG}, and $\tanb$ is the ratio of the higgs vacuum expectation values $v_u/v_d$.  It is important to note that because the PMNS \cite{PMNS} mixing angles are large, the charged lepton mixing angles are expected to be of $\mathcal{O}(1)$.  This means that despite the assumption of having two dominant LNV parameters at the weak interaction basis, after rotating to a diagonal charged lepton basis there will be many LNV parameters of similar order of magnitude.

\begin{table}
\begin{tabular}{l}
\hline
$\Delta m^2_{21}=7.9^{+0.27}_{-0.28}\cdot 10^{-5}\textrm{eV}^2$\phantom{xx}$|\Delta m^2_{31}|=2.6 \pm 0.2 \cdot 10^{-3}\textrm{eV}^2$ \\
$\textrm{sin}^2\theta_{12}=0.31\pm 0.02$ \phantom{hahaha}$\textrm{sin}^2\theta_{23}=0.47^{+0.08}_{-0.07}$\\
$\textrm{sin}^2\theta_{13}=0^{+0.008}_{-0.0}$ \\
\hline
\end{tabular}
\caption{Neutrino oscillation data obtained in \cite{0704.1800}.}
\label{tab:neutrino_data}
\end{table}

\subsection{Neutrino masses}
Since lepton number is violated, the 3 neutrinos mix with the 4 neutralinos.  The 7 $\times$ 7 neutrino-neutralino mass matrix $\mM_N$ can be written as
\eqa
\mM_N &=& \left( \begin{array}{cc}
\mM_{\chi^0} & m^T \\
m & m_{\nu} \end{array} \right),
\qea
where $m_{\nu}$ and $m$ contain the lepton number violating contributions, and $\mM_{\chi^0}$ is the neutralino mass matrix.  An effective 3 $\times$ 3 neutrino mass matrix can be obtained by an `electroweak' see-saw mechanism, in which the neutralinos act as the see-saw.  It is given by
\eqa \label{eq:seesaw}
\mM^{\nu}_{\textrm{eff}} &=& m_{\nu} - m \mM_{\chi^0}^{-1} m^T.
\qea

\subsection{Renormalization and radiative effects}
\begin{figure}
\scalebox{0.8}{
  \begin{picture}(210,75)(0,0)
    \ArrowLine(10,35)(60,35)
    \ArrowArc(80,35)(20,0,180)
    \ArrowArcn(80,35)(20,0,180)
    \ArrowLine(100,35)(150,35)
    \ArrowLine(200,35)(150,35)
    \Vertex(60,35){1.5}
    \Vertex(100,35){1.5}
    \Vertex(150,35){1.5} 
    \put(35,40){$L_i$}
    \put(45,25){$\lambda'_{iaq}$}
    \put(80,60){$Q_a$}
    \put(80,5){$D^c_q$}
    \put(101,24){$(Y_D^*)_{aq}$}
    \put(120,40){$H_D$}
    \put(150,25){$\mu$}
    \put(175,40){$H_U$}
  \end{picture}
}\caption{Dynamical generation of $\mu_i$.}\label{fig:dyn_gen_kap}
\end{figure}
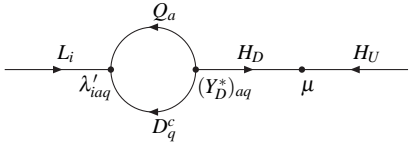

Note that at tree level, $m_{\nu}=0_{3\times 3}$.  The effective mass matrix is proportional to \cite{9408224_9508271}
\eqa
(\mM^{\nu}_{\textrm{eff}})_{ij} &\propto& \Lambda_i\Lambda_j, \nonumber \\
\Lambda_i &\equiv& \mu v_i - v_d \mu_i, \qquad i=\{e,\mu,\tau\},
\qea
so $\mM^{\nu}_{\textrm{eff}}$ depends only on the sneutrino vevs $v_i$ and $\mu_i$ bilinear parameters.  Even though they are not present at $\MX$, they will be generated upon renormalization to lower scales.  A pseudo Feynman diagram for the generation of $\mu_i$ is shown in Fig.~\ref{fig:dyn_gen_kap}.

\begin{figure}
\scalebox{0.8}{
  \begin{picture}(210,75)(0,0)
    \DashArrowArc(105,40)(20,90,180){3}
    \DashArrowArcn(105,40)(20,90,0){3}
    \ArrowArcn(105,40)(20,270,180)
    \ArrowArc(105,40)(20,270,360)
    \ArrowLine(45,40)(85,40)
    \ArrowLine(165,40)(125,40)
    \Vertex(105,20){1.5}
    \Vertex(105,60){1.5}
    \put(35,39){$\nu_i$}
    \put(167,39){$\nu_i$}
    \put(60,28){$\lambda_{iL_jR_k}$}
    \put(128,28){$\lambda_{iL_kR_j}$}
    \put(98,65){$(\mM^{2*}_{\tilde{f}})_{L_kR_k}$}
    \put(100,10){$m^*_{f_j}$}
  \end{picture}
}\caption{A mass insertion diagram which represents the radiative correction to $m_{\nu}$ by $\lambda_{iL_jR_k}$ and $\lambda_{iL_kR_j}$.}\label{fig:SigmaD_MIA}
\end{figure}
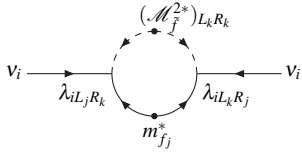

The rank one structure of $\mM^{\nu}_{\textrm{eff}}$ means there is only one massive neutrino at tree level.  The inclusion of radiative corrections is necessary to provide a second mass scale required by the measured neutrino mass squares differences.  The one loop corrections to $(\mM_N)_{ij}$ \cite{Davidson:2000ne} is given by $\frac{1}{2}(\delta\mM_N + \delta\mM_N^T)_{ij}$, where
\eqa \label{eq:one_loop}
(\delta \mM_N)_{ij} &=& (\Sigma_D)_{ij} - (\mM_N)_{ik}(\Sigma_L)_{kj},
\qea
and $\Sigma_D$ and $\Sigma_L$ are mass corrections and wavefunction renormalization respectively.  A mass insertion diagram for the radiative correction of $m_{\nu}$ is shown in Fig.~\ref{fig:SigmaD_MIA}.

While the bilinear parameters which control the tree level mass scale are generated via renormalization group running, the trilinear parameters already present at $\MX$ contribute to the radiative correction of $\mM_N$.  Due to the large logarithmic factor coming from renormalization from $\MX$ to $\sim 1$ TeV, the tree level contribution turns out to be dominant.  The ratio $m^{tree}_{\nu}/m^{loop}_{\nu}$ can be approximated to be \cite{Allanach:2007qc}
\eqa
\label{eq:tree_loop_approx}
\hspace*{-1.0cm}\frac{m^{tree}_{\nu}}{m^{loop}_{\nu}} &\simeq&
-n_{c}\frac{\alpha_{GUT}\textrm{ln}^2(\MX/M_{Z})}{10\pi M_{1/2}(A_0 -
  \mu\tanb)} \nonumber \\
&&\times\frac{(\mM^2_{\tilde{f}})_{L_kL_k}-(\mM^2_{\tilde{f}})_{R_kR_k}}{\textrm{ln}((\mM^2_{\tilde{f}})_{L_kL_k}/(\mM^2_{\tilde{f}})_{R_kR_k})}f^2,
\qea
where $n_c$ is 1 for $\lam_{ijk}$ and 3 for $\lam'_{ijk}$, $f$ is a dimensionless function of $\mathcal{O}(10)$\cite{0309196}, and $\mM^2_{\tilde{f}}$ is mass matrix of scalar $\tilde{f}$.  A rough scan in the mSUGRA parameter space gives ratios of $\mathcal{O}(30-200)$.  On the other hand, consistency with neutrino oscillation data requires the mass ratio of the two heaviest neutrinos to be at most of $\mathcal{O}(5)$.  This implies for our simple model to work, the combined contribution to the dynamical generation of the bilinear parameters must be suppressed to decrease the mass ratio described above.  The charged lepton mixing angles also need to be fixed.  The best fit parameters are obtained by a numerical procedure discussed next.

\section{Numerical procedure}

We restrict ourselves to the SPS1a \cite{SPS1a} point
\begin{equation}
\begin{array}{l}
\Mhalf=250\textrm{GeV},\phantom{xx}\MO=100\textrm{GeV},\phantom{xx}\AO=-100\textrm{GeV},\\
\sgnmu = +1, \quad\quad\quad\tanb =10,
\end{array}
\end{equation}
and vary only the 2 LNV parameters and the 3 charged lepton mixing angles $\theta^{l}$.  After specifying these parameters at $\MX$, the Lagrangian is rotated to a diagonal charged lepton basis using the $\theta^{l}$'s.  A modified version of $\softsusy$ \cite{softsusy} code, including the full set of 1-loop LNV MSSM RGEs contributions, is used to run to $\sim\mathcal{O}(500)$GeV at which the EWSB conditions are imposed, and the neutrino mass matrix is calculated.  All 1-loop corrections to the latter are calculated numerically except for the CP-even and CP-odd neutral scalar contributions, which are computed in an analytic expansion in the relevant LNV parameters to avoid fluctuations from strong numerical cancellations.   The optimization of the best fit parameters is performed using MINUIT with the neutrino oscillation data displayed in Table \ref{tab:neutrino_data}.

In Table \ref{tab:result}, a selected set of best fit parameters is displayed.  We note that no attempt to exhaust all combinations of $\lam$ and $\lam'$ couplings is made.  There are other parameter sets which fits the neutrino oscillation data, but are not presented here since they violate other experimental constraints \cite{EWconstraints}, such as leptonic FCNCs.

\begin{table}
  \centering
  \scalebox{0.68}{
    \begin{tabular}{lllllll}
      \hline
      Normal hierarchy&&&&&&\\
      \hline
      $\Lambda_1$&$\Lambda_2$&$\theta^l_{12}$&$\theta^l_{13}$&$\theta^l_{23}$&$\Delta_{FT}$&$\chi^2$ \\
      \hline
      $\lambda'_{233}=-2.50\cdot 10^{-6}$ &$\lambda_{233}=4.07\cdot 10^{-5}$& $0.460$ & $0.389$& $0.305$ & $8.09$ &-\\
      $\lambda'_{233}=-2.50\cdot 10^{-6}$ &$\lambda_{211}=4.07\cdot 10^{-5}$& $1.990$ & $1.082$& $0.632$ & $8.10$ &-\\
      $\lambda'_{233}=-3.41\cdot 10^{-6}$ &$\lambda_{321}=9.87\cdot 10^{-5}$& $0.448$ & $0.400$& $2.891$ & $12.4$ &-\\ \hline
      $\lambda'_{122}=-1.14\cdot 10^{-4}$ &$\lambda_{122}=4.06\cdot 10^{-5}$& $1.193$ & $0.191$& $1.174$ & $11.8$ &-\\
      $\lambda'_{122}=-8.98\cdot 10^{-5}$ &$\lambda_{123}=1.03\cdot 10^{-4}$& $2.107$ & $0.175$& $1.181$ & $9.44$ &-\\
      $\lambda'_{122}=-8.60\cdot 10^{-5}$ &$\lambda_{133}=4.10\cdot 10^{-5}$& $0.998$ & $0.282$& $0.418$ & $8.00$ &-\\\hline
      Inverted hierarchy&&&&&&\\
      \hline
      $\Lambda_1$&$\Lambda_2$&$\theta^l_{12}$&$\theta^l_{13}$&$\theta^l_{23}$&$\Delta_{FT}$&$\chi^2$ \\
      \hline
      $\lambda'_{233}=-5.69\cdot 10^{-6}$ &$\lambda_{233}=1.36\cdot 10^{-4}$& $1.558$ &$0.815$ &$0.146$&$755$& $0.01$\\
      $\lambda'_{233}=-5.69\cdot 10^{-6}$ &$\lambda_{211}=1.32\cdot 10^{-4}$& $1.388$ &$0.760$ &$0.141$&$758$& $0.06$\\
      $\lambda'_{233}=-5.68\cdot 10^{-6}$ &$\lambda_{123}=1.43\cdot 10^{-4}$& $1.814$ &$-0.758$&$0.142$&$726$& $0.05$\\ \hline
      $\lambda'_{122}=-1.96\cdot 10^{-4}$ &$\lambda_{122}=1.25\cdot 10^{-4}$& $0.135$ &$0.102$ &$0.798$&$988$& $0.43$\\
      $\lambda'_{122}=-1.94\cdot 10^{-4}$ &$\lambda_{132}=1.47\cdot 10^{-4}$& $3.032$ &$0.087$ &$0.932$&$743$& $2.85$\\
      $\lambda'_{122}=-1.96\cdot 10^{-4}$ &$\lambda_{123}=1.46\cdot 10^{-4}$& $0.144$ &$0.094$ &$0.690$&$736$& $0.52$\\\hline
    \end{tabular}
  }\caption{Set of best fit parameters at SPS1a, assuming both the normal and inverted hierarchies.  See text for more discussions.}\label{tab:result}
\end{table}

In the table, we display values of $\chi^2$ and fine tuning measure $\Delta_{FT}$ of the best fit points.  The tuning is much more severe for an inverted hierarchy because in the former case, instead of having a symmetry to relate the two neutrino masses to be (almost) equal, the LNV parameters need to be arranged to make the tree level mass scale quasi-degenerate with the scale of radiative corrections in $\mM^{\nu}_{\textrm{eff}}$.  The lack of an underlying flavour symmetry also means the observed near tri-bi maximal mixing \cite{Harrison} should be regarded as accidental.

A few general observations on possible LHC signatures are in order.  The LNV parameters are typically too small to affect decay of the sparticles produced, except for the LSP.  On the other hand, the $\mathcal{O}(1)$ charged lepton mixing angles means that the LSP is expected to decay in many different channels with similar branching ratios.  For the SPS1a spectrum we examine, the LSP can decay with a displaced vertex of $\mathcal{O}(0.1)$mm.  Such displacements should not be immediately obvious, but may be searched for to provide support for this type of models.  However if the LSP is a stau for instance, it should undergo 2 body decay at the interaction point.

\section{Summary}
We have discussed SUSY with lepton number violation as an alternative to neutrino masses.  In high scale constructions, renormalization effect typically leads to neutrino masses incompatible with neutrino oscillation data, unless the trilinear LNV parameters are arranged to suppress the dynamical generation of the bilinear LNV parameters responsible for the tree level $m_{\nu}$.  We performed a numerical study to fit the oscillation data in an mSUGRA model, assuming two dominant trilinear LNV couplings defined in a weak interaction basis, and discuss collider effects that could be observed at the LHC.

%%%%%%%%%%%%%%%%%%%%%%%%%%%%%%%%%%%%%%%%%%%%%%%%
%% BACKMATTER
%%%%%%%%%%%%%%%%%%%%%%%%%%%%%%%%%%%%%%%%%%%%%%%%

\begin{theacknowledgments}
This work has been partially supported by STFC.  CHK is funded by a Hutchison Whampoa Dorothy Hodgkin Postgraduate Award.  We thank the members of the Cambridge SUSY working group and S Rimmer for useful conversations.  The computational work has been performed using the Cambridge eScience CAMGRID computing facility, with the invaluable help of Mark Calleja.
\end{theacknowledgments}

\bibliographystyle{aipproc}

\end{document}